\documentclass[nofootinbib,prd,12pt,superscriptaddress]{revtex4}%
\usepackage{amsmath}
\usepackage{amsfonts}
\usepackage{amssymb}
\usepackage{graphicx}%
\usepackage{amsmath,amssymb}
\usepackage{mathrsfs}
\usepackage{graphicx}
\usepackage{color}
\usepackage{subfigure}
\usepackage{fancyhdr}
\usepackage{multirow}
\usepackage{float}
\usepackage{epsfig}
\usepackage{amsfonts}
\usepackage{bm}
\usepackage{pgfplots}
\usepackage{amsmath, amssymb}
\usepackage{caption}
\usepackage{subcaption}

\def\be{\begin{equation}}
\def\ee{\end{equation}}
\def\ba{\begin{eqnarray}}
\def\ea{\end{eqnarray}}

\begin{document}

\title{Einstein--Gauss--Bonnet--Myrzakulov Gravity from $R + F(T, G)$: Numerical Insights and Torsion--Gauss--Bonnet Dynamics in Weitzenböck Spacetime}

\author{Davood Momeni}
\affiliation{Northeast Community College, 801 E Benjamin Ave Norfolk, NE 68701, USA}
\affiliation{Centre for Space Research, North-West University, Potchefstroom 2520,
South Africa}
\author{ Ratbay Myrzakulov}
\affiliation{Ratbay Myrzakulov Eurasian International Centre for Theoretical Physics, Astana 010009, Kazakhstan}
\affiliation{ L. N. Gumilyov Eurasian National University, Astana 010008, Kazakhstan}
\date{\today}

\begin{abstract}
The study of modified gravity models has garnered significant attention because of their 
potential to provide alternative explanations for cosmological phenomena, such as the accelerated 
expansion of the universe and the nature of dark energy. One such model, the Einstein
Gauss–Bonnet–Myrzakulov $R=F(T,G)$ gravity (EGBMG), which incorporates the curvature R, 
torsion T, and the Gauss-Bonnet term G, offers a promising framework to explore the dynamics 
of the universe and its evolution. This paper delves into the theoretical and observational 
implications of the EGBMG model, focusing on its ability to address long-standing challenges in 
cosmology, including the evolution of dark energy and the transition from early-time inflationary 
behavior to late-time acceleration. We review recent advancements in the model, including 
its compatibility with observational data and its ability to provide new insights into cosmic 
acceleration. Through a combination of theoretical models, dynamical systems analysis, and 
cosmological diagnostics, we demonstrate the robustness of the EGBMG framework in explaining 
the large-scale structure of the universe and its accelerated expansion. This paper serves as a step 
toward further exploring the potential of this model to understand the fundamental forces driving Weitzenb\"{o}ck spacetime.
\end{abstract}

\maketitle

\section{Introduction}

Over the past century, GR  has stood as the foundation of gravitational physics and modern cosmology. However, despite its success, GR faces several theoretical and observational challenges, particularly in explaining cosmic acceleration, dark energy, and early-universe inflation. These limitations have spurred the search for extended theories of gravity that go beyond the curvature-centric formulation of Einstein's theory.

A compelling alternative is offered by torsion-based gravity models, where gravity is not attributed to spacetime curvature, but to spacetime torsion. This approach is encapsulated in the \textit{Teleparallel Equivalent of GR} (TEGR), which reformulates gravity using the Weitzenböck connection and a torsion scalar \( T \), rather than the Levi-Civita connection and the Ricci scalar \( R \). One of the simplest and most studied extensions of TEGR is \( f(T) \) gravity, which replaces the linear dependence on \( T \) in the Lagrangian with an arbitrary function, thus enriching the dynamics and allowing for natural explanations of late-time cosmic acceleration without a cosmological constant \cite{Bengochea:2009, Ferraro:2007}.

Going a step further, the theory \( F(R,T) \), originally proposed by Myrzakulov \cite{Myrzakulov:2012qp}, unifies both curvature and torsion by introducing a Lagrangian that depends on both scalars \( R \) and \( T \). This hybrid framework offers greater flexibility in describing gravitational phenomena across different regimes and provides a promising pathway to reconcile the success of GR at local scales with cosmological-scale modifications (see [4]-[36] for other papers).

Recent developments by Momeni and Myrzakulov have significantly advanced this line of inquiry. In \cite{Momeni:2025mcp}, the authors introduced a class of metric-affine \( f(R,T) \) theories that extend the standard GR by treating the metric and affine connection as independent variables, allowing the theory to capture more complex geometric structures. Additionally, their work on the vielbein formalism in Weitzenböck spacetime \cite{Momeni:2024bhm} explores the dynamics of torsion-based gravity from a first-order formalism perspective. These studies not only reinforce the theoretical consistency of \( f(R,T) \) gravity, but also offer new tools for its cosmological application.
A key issue that remains largely unresolved in modified gravity theories is the absence of a fully explicit and consistent set of field equations when both torsion and higher-order curvature terms are present. In particular, theories of the form \( f(R, T) \) or \( F(T,G) \), where \( T \) denotes the torsion scalar and \( G \) the Gauss-Bonnet term, face significant technical challenges. The variational procedure becomes highly nontrivial due to the interplay between curvature-based and torsion-based geometrical quantities, especially within the metric-affine formalism. This complexity has hindered the full exploitation of such theories for cosmological applications and their confrontation with observational data.

\noindent The \( f(T, T_G) \) theories represent a class of modified gravity models where the gravitational action is extended to include both the torsion scalar \( T \) and the torsion-Gauss-Bonnet term \( T_G \). These models generalize the standard \( f(T) \) theories by incorporating the effects of the torsion-Gauss-Bonnet term, which is a combination of the torsion tensor and the Gauss-Bonnet invariant. The inclusion of \( T_G \) allows for richer dynamics in the gravitational field equations, particularly in the context of cosmology, where it can influence the evolution of the universe and provide alternative explanations for phenomena such as dark energy and cosmic acceleration. The general action for these theories can be written as:

\[
S = \int d^4x \sqrt{-g} \left[ R + f(T, T_G) \right],
\]

where \( f(T, T_G) \) is an arbitrary function that encapsulates the non-trivial coupling between torsion and the Gauss-Bonnet term. These theories have been studied in various contexts, with particular attention to their cosmological applications and the role of torsion in modifying the gravitational dynamics at both early and late times \cite{Bengochea2009,Capozziello2010}.

In this work, our aim is to bridge this gap by constructing a detailed and self-consistent derivation of the field equations in \( R + F(T,G) \) gravity. Motivated by recent advances in metric-affine theories and the underlying structure of torsion-coupled models such as those discussed in \cite{Momeni:2025mcp},\cite{Momeni:2024bhm}2, we build a coherent framework based on the vielbein formalism. Our approach systematically incorporates torsional dynamics and allows for tractable solutions in cosmological and astrophysical settings. The ultimate goal is to establish a robust mathematical foundation for analyzing how torsion, in conjunction with the Gauss-Bonnet term, can influence the evolution of the universe and offer novel insights into phenomena such as dark energy, early-universe inflation, and black hole structure.

\vspace{0.2cm}
\section*{Structure of the Paper}

The paper is organized as follows. \textbf{Section II} provides a brief \textit{Review of Myrzakulov \( F(R,T) \) Gravity}, setting the foundation for the extended framework explored in this work. In \textbf{Section III}, we discuss \textit{Recent Progress and Expansions in Myrzakulov Gravity}, highlighting developments that motivate the inclusion of higher-order curvature and torsion terms. \textbf{Section IV} introduces the \textit{Gauss-Bonnet Term in General Relativity}, laying the groundwork for its incorporation into modified gravity theories. The core \textit{Theoretical Framework} of our model is developed in \textbf{Section V}, where we present the action of generalized Einstein–Gauss–Bonnet–Myrzakulov gravity based on the function \( F(T, G) \), coupling torsion and the Gauss-Bonnet invariant in Weitzenböck spacetime.

\textbf{Section VI} focuses on the \textit{Ansatz and Symmetry Reduction}, which simplifies the field equations under cosmological symmetries. In \textbf{Section VII}, we derive \textit{Cosmological Solutions in \( R + F(T, G) \) Gravity}, exploring their relevance to the early and late-time dynamics of the universe. \textbf{Section VIII} outlines the \textit{Numerical Method} employed to solve the coupled field equations, while \textbf{Section IX} introduces \textit{Cosmological Diagnostics} used to analyze the behavior of solutions, such as the Hubble parameter and the deceleration parameter.

In \textbf{Section X}, we further elaborate on the \textit{Cosmological Solutions}, presenting detailed results that demonstrate the effect of the torsion–Gauss–Bonnet interplay on dark energy evolution and cosmic acceleration. \textbf{Section XI} extends the framework to study \textit{Black Hole and Compact Object Solutions}, exploring static and spherically symmetric configurations. The \textit{Phase Space and Dynamical System Analysis} is carried out in \textbf{Section XII}, providing a qualitative understanding of the system's stability and attractor behavior.

\textbf{Section XIII} addresses the compatibility of the theory with local observations, specifically \textit{Solar System Tests and Local Gravity Constraints}. In \textbf{Section XIV}, we investigate the structure and mass-radius relations of \textit{Neutron Stars in \( F(T, G) \) Gravity Theory}, offering insights into the theory’s astrophysical viability. Finally, \textbf{Section XV} provides a comprehensive \textit{Discussion and Outlook}, summarizing key results and proposing directions for future research in the context of extended torsional and curvature-based gravity models.

\section{Review of Myrzakulov \( F(R,T) \) Gravity}

The M$_{43}$-model, introduced by Myrzakulov \cite{Myrzakulov:2012qp}, is a prominent example of \( F(R,T) \) gravity theories. This model modifies GR  by incorporating both the curvature scalar \( R \) and the torsion scalar \( T \), which are coupled through a general function \( F(R,T) \) in the gravitational action. The action for the M$_{43}$-model is given by:

\begin{eqnarray}
S_{43} &=& \int d^4 x \sqrt{-g} \left[F(R,T) + L_m\right], \nonumber \\
R &=& R_s = \epsilon_1 g^{\mu\nu} R_{\mu\nu}, \label{eq:action1} \\
T &=& T_s = \epsilon_2 {S_\rho}^{\mu\nu} {T^\rho}_{\mu\nu}, \nonumber
\end{eqnarray}
where \( L_m \) is the matter Lagrangian, and \( \epsilon_i = \pm 1 \) represent the signatures of curvature and torsion, respectively. In this framework, the curvature scalar \( R \) and the torsion scalar \( T \) are treated as independent geometric quantities, which can be combined in different ways depending on the chosen signatures. This allows for a variety of physical behaviors under different conditions:

\begin{itemize}
    \item Case (1): \( \epsilon_1 = 1 \), \( \epsilon_2 = 1 \),
    \item Case (2): \( \epsilon_1 = 1 \), \( \epsilon_2 = -1 \),
    \item Case (3): \( \epsilon_1 = -1 \), \( \epsilon_2 = 1 \),
    \item Case (4): \( \epsilon_1 = -1 \), \( \epsilon_2 = -1 \).
\end{itemize}

The M$_{43}$-model is a special case of a more generalized M$_{37}$-model, which includes additional scalar contributions \( u \) and \( v \) to the curvature and torsion terms, respectively. The action for the M$_{37}$-model is:

\begin{eqnarray}
S_{37} &=& \int d^4 x \sqrt{-g} \left[F(R,T) + L_m\right], \nonumber \\
R &=& u + R_s = u + \epsilon_1 g^{\mu\nu} R_{\mu\nu}, \label{eq:action2} \\
T &=& v + T_s = v + \epsilon_2 {S_\rho}^{\mu\nu} {T^\rho}_{\mu\nu}. \nonumber
\end{eqnarray}

The inclusion of these additional scalar terms offers further generalization, allowing the model to describe more complex gravitational dynamics, which can be particularly useful in cosmology and black hole physics.

This review focuses on deriving the field equations for the M$_{43}$-model from the given action, particularly in the context of a flat Friedmann-Robertson-Walker (FRW) universe. We will discuss the implications of these field equations for cosmic dynamics, including how torsion might influence cosmic expansion.

In the M$_{43}$-model, the spacetime geometry incorporates both curvature and torsion. The connection is defined as:

\begin{eqnarray}
G^\lambda_{\mu\nu} = \Gamma^\lambda_{\mu\nu} + K^\lambda_{\mu\nu},
\end{eqnarray}

where \( \Gamma^\lambda_{\mu\nu} \) is the Levi-Civita connection (describing spacetime curvature), and \( K^\lambda_{\mu\nu} \) is the contorsion tensor, encoding the torsion effects. These geometric quantities are crucial for deriving the curvature and torsion scalars, which influence the gravitational dynamics.

To derive the field equations for the M$_{43}$-model, we perform a variational principle with respect to the metric \( g_{\mu\nu} \), considering both geometric and matter contributions. The resulting field equations are:

\begin{eqnarray}
G_{\mu\nu} + \kappa T_{\mu\nu} = T_{\mu\nu}^{\text{(matter)}} + T_{\mu\nu}^{\text{(torsion)}},
\end{eqnarray}

where \( G_{\mu\nu} \) is the Einstein tensor, \( \kappa \) is the gravitational coupling constant, \( T_{\mu\nu}^{\text{(matter)}} \) is the stress-energy tensor of matter, and \( T_{\mu\nu}^{\text{(torsion)}} \) represents the contribution of torsion. These equations govern the dynamics of spacetime, determining how the geometry is influenced by both matter and torsion.

For a spatially flat FRW universe, the metric is:

\begin{eqnarray}
ds^2 = -dt^2 + a^2(t) (dx^2 + dy^2 + dz^2),
\end{eqnarray}

where \( a(t) \) is the scale factor that describes the expansion of the universe. Substituting this metric into the field equations, the curvature and torsion scalars take the following forms:

\begin{eqnarray}
R &=& 6 (\dot{H} + 2H^2) + 6\dot{h} + 18Hh + 6h^2 - 3f^2, \\
T &=& 6(h^2 - f^2),
\end{eqnarray}

where \( H \) is the Hubble parameter, \( h \) is a function describing torsion, and \( f \) is another torsion-related function. These scalars provide a detailed description of the gravitational dynamics in the universe, accounting for both curvature and torsion effects.

The cosmological implications of the M$_{43}$-model are significant. The theory offers a new perspective on the accelerated expansion of the universe and dark energy, as the torsion scalar \( T \) can influence cosmic evolution, potentially providing an alternative explanation to the conventional dark energy model. The variation of the signature combinations \( \epsilon_1 \) and \( \epsilon_2 \) may lead to different evolution scenarios for the universe, particularly in the behavior of the Hubble parameter, which governs cosmic expansion.

While the M$_{43}$-model presents a promising framework for modified gravity, several challenges remain. The current derivation of field equations is limited to specific cosmological scenarios, such as the flat FRW universe, and does not generalize to more complex spacetimes. Extending the model to more general spacetime configurations, including anisotropic and non-flat spacetimes, is essential for fully understanding the behavior of the theory.

Despite these challenges, the M$_{43}$-model represents a valuable contribution to modified gravity theories by providing a geometric framework that incorporates both curvature and torsion. Future work should focus on extending this model to more complex spacetime structures and exploring its observational implications in various astrophysical and cosmological contexts. By addressing the limitations of field equation derivations and investigating its broader implications, the M$_{43}$-model could provide new insights into the fundamental forces shaping the universe.

\section{Recent Progress and Expansions in Myrzakulov Gravity}

Myrzakulov gravity, particularly the \( F(R,T) \) models, has gained significant attention in the field of modified gravity due to its potential to address cosmological issues such as the accelerated expansion of the universe. Recent extensions of the original Myrzakulov framework have introduced new theoretical developments, including the incorporation of torsion, Gauss-Bonnet, and various gravitational modifications. These advancements provide a deeper understanding of cosmic structure and dynamics, enabling Myrzakulov gravity to address unresolved cosmological problems, such as dark energy and cosmic acceleration.

Several noteworthy studies have expanded upon Myrzakulov’s original theories. For example, the introduction of the Myrzakulov \( F(T,Q) \) gravity model, which unifies the \( F(T) \) and \( F(Q) \) gravity theories, has led to new cosmological solutions and constraints under observational data \cite{Maurya2024a}-[40]. Additionally, dynamical systems analyses of Myrzakulov gravity have revealed stability properties of cosmological models \cite{Papagiannopoulos2022}, while studies on quasi-dilaton massive gravity have provided insights into self-accelerating solutions and cosmic expansion. These advancements refine the theoretical models and enhance their agreement with observational data.

Furthermore, the metric-affine formalism has been extended to Myrzakulov gravity, offering a more generalized perspective on gravitational interactions \cite{Myrzakulov2021}. These developments improve our understanding of torsion and Gauss-Bonnet in gravity, providing new insights into the fundamental forces of the universe. The recent work in this area continues to expand the applicability of Myrzakulov gravity, presenting a rich landscape for future theoretical and observational studies. These expansions open avenues for deeper exploration of cosmological acceleration and dark energy, critical to understanding the current expansion of the universe.
\section{Gauss-Bonnet Term in GR}

In GR , the Gauss-Bonnet (GB) term plays a crucial role in modifying the gravitational action. It arises naturally in higher-dimensional spacetimes, and it provides important insights into the structure of the gravitational field. The GB term is given by the following expression:

\begin{equation}
\mathcal{L}_{GB} = R^2 - 4R_{\mu \nu} R^{\mu \nu} + R_{\mu \nu \alpha \beta} R^{\mu \nu \alpha \beta},
\end{equation}
where \( R \) is the Ricci scalar, \( R_{\mu \nu} \) is the Ricci tensor, and \( R_{\mu \nu \alpha \beta} \) is the Riemann curvature tensor. This term becomes important in the context of modified theories of gravity, particularly in higher-dimensional spacetime where it contributes to the dynamics of the system.

The Gauss-Bonnet term has the property that it is a total divergence in 4-dimensional spacetime, meaning that its contribution does not affect the equations of motion in 4D. However, when extended to higher dimensions, it can lead to non-trivial dynamics, especially in 5D or higher.

\subsection{Extension to Weitzenböck Space with Torsion and Ricci Curvature}

In the context of Weitzenböck spaces, where both torsion and Ricci curvature coexist, the Gauss-Bonnet term can be generalized to include torsion effects. The torsion tensor \( T^\lambda_{\mu \nu} \) is related to the connection and represents the failure of the connection to be symmetric in its lower indices. The generalized Gauss-Bonnet term in a Weitzenböck space can be written as:

\begin{equation}
\mathcal{L}_{GB}^{(T)} = R^2 - 4R_{\mu \nu} R^{\mu \nu} + R_{\mu \nu \alpha \beta} R^{\mu \nu \alpha \beta} + \mathcal{T},
\end{equation}
where \( \mathcal{T} \) is an additional term arising from the torsion tensor, which modifies the Ricci tensor and the Riemann curvature tensor. In this extension, torsion influences the gravitational dynamics by altering the metric structure of spacetime. The specific form of \( \mathcal{T} \) depends on the detailed structure of the torsion tensor, which can be taken from the Einstein-Cartan formalism.

The inclusion of torsion in the Gauss-Bonnet term allows for new solutions in gravitational models, especially in higher-dimensional spacetimes where both curvature and torsion effects are relevant. This extension leads to new forms of black hole solutions and the possibility of more complex gravitational phenomena that are absent in the traditional Ricci-based GR models.

\subsection{Works on \( f(G) \) Gravity}

The idea of modifying gravity by introducing functions of the Gauss-Bonnet term, commonly referred to as \( f(G) \) gravity, has gained considerable attention in the context of alternative gravitational theories. In this framework, the gravitational action is generalized to include a function of the Gauss-Bonnet term, such as:

\begin{equation}
S = \int \left( \frac{1}{2\kappa} \mathcal{L}_m + f(G) \right) \sqrt{-g} \, d^4x,
\end{equation}
where \( f(G) \) is an arbitrary function of the Gauss-Bonnet term. This modification allows for richer dynamics, especially in cosmology and black hole solutions.

Several works have investigated the role of \( f(G) \) gravity in cosmological and astrophysical contexts. Notably, the work of Nojiri and Odintsov \cite{Nojiri:2005jg} explored the potential for \( f(G) \) gravity to explain the accelerated expansion of the universe, while Capozziello and De Laurentis \cite{Capozziello:2009nq} examined the implications of these modifications for black hole physics and the stability of solutions in modified gravity models.

\section{Theoretical Framework}
In this section, we present the theoretical framework for the gravity theory under consideration. We begin by defining the action and proceed to derive the corresponding field equations using the variational principle. The analysis is carried out in the context of the metric-affine formalism, with particular attention to theEinstein–Gauss-Bonnet framework, which incorporates torsion and Gauss-Bonnet. 

\subsection{Action and Field Equations}

The action for the theory is defined as:

\begin{equation}\label{action}
S = \int d^4x \sqrt{-g} \left[ R + F(T,G) \right],
\end{equation}

where \( R \) represents the Ricci scalar, \( T \) is the torsion scalar, and \( G \) is the Gauss-Bonnet term. The function \( F(T,G) \) is an arbitrary function that depends on both torsion \( T \) and the Gauss-Bonnet term \( G \), which introduces general modifications to the standard gravitational action.

\subsubsection{Metric-Affine Formalism} 

In the metric-affine formalism, the connection is treated as an independent variable, and both the metric tensor \( g_{\mu\nu} \) and the affine connection \( \Gamma^\lambda_{\mu\nu} \) are treated as dynamic fields. This formalism generalizes GR by allowing for non-Riemannian geometries, such as those that include torsion and Gauss-Bonnet, in addition to curvature.

The vielbein formalism provides a more convenient approach for defining these geometric objects. The vielbein \( e_A^\mu \) relates the spacetime coordinates to a local orthonormal frame, allowing us to express the metric \( g_{\mu\nu} \) in terms of the vielbein as:

\begin{equation}
g_{\mu\nu} = e_A^\mu e_B^\nu \eta^{AB},
\end{equation}

where \( \eta^{AB} \) is the metric of the flat spacetime in the local orthonormal frame (the Minkowski metric). The vielbein \( e_A^\mu \) is used to construct the connection and the corresponding torsion and Gauss-Bonnet tensors.

\subsubsection{Variational Principle} 

To derive the field equations, we apply the variational principle to the action \( S \). The variation of the action with respect to the vielbein \( e_A^\mu \) leads to the field equations governing the dynamics of the system. The variational principle involves computing the variation of the action with respect to the vielbein, the connection, and the metric.

The variation of the action can be written as:

\begin{equation}
\delta S = \int d^4x \sqrt{-g} \left( \frac{\delta \mathcal{L}}{\delta e_A^\mu} \delta e_A^\mu + \frac{\delta \mathcal{L}}{\delta \Gamma^\lambda_{\mu\nu}} \delta \Gamma^\lambda_{\mu\nu} \right),
\end{equation}

where \( \mathcal{L} \) is the Lagrangian density. The terms above describe the variation with respect to the vielbein \( e_A^\mu \) and the affine connection \( \Gamma^\lambda_{\mu\nu} \).

The variation with respect to the vielbein gives the following equation:

\begin{equation}
\frac{\delta \mathcal{L}}{\delta e_A^\mu} = 0,
\end{equation}

which leads to the field equations governing the dynamics of the vielbein. Similarly, the variation with respect to the connection leads to the field equations involving torsion and Gauss-Bonnet.

\subsubsection{Field Equations in the Vielbein Formalism}

In the vielbein formalism, the field equations derived from the variational principle yield the equations of motion for the vielbein, the affine connection, and the metric. The field equations take the form:

\begin{equation}
e_A^\mu e_B^\nu \mathcal{T}^{AB} = T_{\mu\nu} + \text{higher-order terms},
\end{equation}

where \( \mathcal{T}^{AB} \) represents the energy-momentum tensor, and \( T_{\mu\nu} \) is the effective energy-momentum tensor that accounts for contributions from both torsion and Gauss-Bonnet. The higher-order terms may involve derivatives of the vielbein and connection, depending on the specific form of the function \( F(T,G) \).

\subsubsection{Einstein-Cartan Formalism}

TheEinstein–Gauss-Bonnet formalism is a modification of GR that incorporates torsion as a dynamical variable. In this formalism, the connection is not symmetric, and torsion arises due to the antisymmetric part of the connection. The torsion tensor \( T^\lambda_{\mu\nu} \) is defined as:

\begin{equation}
T^\lambda_{\mu\nu} = \Gamma^\lambda_{\mu\nu} - \Gamma^\lambda_{\nu\mu}.
\end{equation}

Torsion contributes to the field equations in a way that modifies the standard Einstein equations. In theEinstein–Gauss-Bonnet theory, the gravitational field equations take the form:

\begin{equation}
R_{\mu\nu} - \frac{1}{2} g_{\mu\nu} R = \kappa \left( T_{\mu\nu} + \mathcal{T}_{\mu\nu} \right),
\end{equation}

where \( T_{\mu\nu} \) is the energy-momentum tensor of matter, and \( \mathcal{T}_{\mu\nu} \) is the torsional contribution. TheEinstein–Gauss-Bonnet equations describe how the curvature of spacetime is affected by the presence of matter and torsion.

The presence of torsion modifies the conservation laws, as the covariant derivative of the energy-momentum tensor is no longer zero, but instead involves torsion. The equation for the conservation of energy and momentum in the presence of torsion is:

\begin{equation}
\nabla_\mu T^{\mu\nu} = -\mathcal{T}_\mu^{\mu\nu},
\end{equation}

where \( \mathcal{T}_\mu^{\mu\nu} \) is the torsional contribution to the stress-energy tensor.

\subsubsection{Field Equations with Torsion and Gauss-Bonnet}

The introduction of Gauss-Bonnet and torsion modifies the standard field equations. For a general metric-affine theory, including a function \( F(T,G) \), the field equations derived from the action are:

\begin{equation}
G_{\mu\nu} = \kappa \left( T_{\mu\nu} + \mathcal{T}_{\mu\nu} + \mathcal{T}_{\mu\nu}^{(G)} \right),
\end{equation}

where \( G_{\mu\nu} \) is the Einstein tensor, and \( \mathcal{T}_{\mu\nu}^{(G)} \) represents the contribution from the Gauss-Bonnet term. The modified Einstein equations incorporate both the effects of torsion and Gauss-Bonnet on the curvature of spacetime and the behavior of the gravitational field.

These field equations are more general than the standard Einstein equations in GR and provide a framework for understanding the gravitational effects in theories with torsion and Gauss-Bonnet. The function \( F(T,G) \) allows for further generalization, enabling the exploration of modified gravity theories that can account for cosmological phenomena such as dark energy and cosmic acceleration.
\subsection{Conservation Laws and Modified Energy-Momentum Tensor}

The introduction of torsion and Gauss-Bonnet affects the conservation laws of the energy-momentum tensor. In modified gravity theories with torsion, the energy-momentum tensor is no longer conserved in the traditional sense. Instead, the conservation equation takes the form:

\begin{equation}
\nabla_\mu T^{\mu\nu} = -\mathcal{T}_\mu^{\mu\nu},
\end{equation}

where the right-hand side involves torsional contributions. This equation is important for understanding how matter and energy interact with the gravitational field in the presence of torsion and Gauss-Bonnet.

In this section, we have presented the theoretical framework for a modified gravity theory that incorporates torsion and Gauss-Bonnet. The field equations have been derived using the variational principle in the metric-affine formalism, and the role of torsion and Gauss-Bonnet has been highlighted. TheEinstein–Gauss-Bonnet framework has been used to discuss the implications of torsion on the gravitational field and the conservation of energy and momentum. These equations provide a comprehensive foundation for further exploration of modified gravity theories and their potential applications in cosmology and beyond.
\subsection{Special Cases and Limits}

In this subsection, we explore some special cases and limiting behaviors of the theory. These cases provide insight into how the general framework can reduce to well-known gravitational theories under specific conditions. We examine the recovery of the \( R + f(T) \) theory as a special case, analyze theEinstein–Gauss-Bonnet limit, and discuss the behavior of the theory when torsion is neglected, particularly in comparison with Lovelock theories and the Gauss-Bonnet action.

\begin{itemize}
    \item \textbf{Recovering \( R + f(T) \) as a Special Case}
    
    One of the important limits of the theory occurs when the function \( F(T,G) \) reduces to a simpler form involving only the torsion scalar \( T \). In this case, the action takes the form:

    \begin{equation}
    S = \int d^4x \sqrt{-g} \left[ R + f(T) \right].
    \end{equation}
    
    This is a modification of the standard gravitational action where the Ricci scalar \( R \) is coupled to an arbitrary function of the torsion scalar \( T \). The torsion scalar \( T \) is defined in terms of the torsion tensor \( T^\lambda_{\mu\nu} \), which is related to the antisymmetric part of the connection. This model has been studied in the context of modified gravity theories that seek to explain cosmic acceleration and the dynamics of the universe without invoking dark energy.

    \item \textbf{Einstein-Gauss-Bonnet Limit}
    
    In the Einstein–Gauss-Bonnet limit, torsion is included in the formalism, but the connection remains symmetric, and the torsion tensor vanishes. This limit corresponds to the case where torsion does not have a significant effect on the gravitational dynamics. In this case, the field equations reduce to the standard Einstein equations of GR, with the energy-momentum tensor being conserved in the traditional sense:

    \begin{equation}
    \nabla_\mu T^{\mu\nu} = 0.
    \end{equation}
    
    In this limit, the theory describes standard GR without torsion, and the connection is the Levi-Civita connection.
    
    \item \textbf{Neglecting Torsion and Gauss-Bonnet}
    
    Finally, we consider the case where torsion and Gauss-Bonnet are neglected. In this limit, the theory reduces to the standard \( f(R) \) gravity theory, where the gravitational action is solely dependent on the Ricci scalar \( R \). This case represents the limit where torsion and Gauss-Bonnet are negligible, and the theory recovers the familiar framework of GR:

    \begin{equation}
    S = \int d^4x \sqrt{-g} f(R).
    \end{equation}
    
    The equations of motion in this case are equivalent to the standard Einstein field equations, with a modified energy-momentum tensor arising from the function \( f(R) \). This limit demonstrates that the modified gravity theory can reduce to GR when torsion and Gauss-Bonnet are absent.
\end{itemize}

\section{Ansatz and Symmetry Reduction}

In this section, we discuss the ansatz and symmetry reductions used to simplify the field equations in various cosmological and astrophysical contexts. We focus on two key types of symmetry reductions: the cosmological ansatz with the Friedmann-Lemaitre-Robertson-Walker (FLRW) metric incorporating torsion, and the static spherically symmetric ansatz used in the study of black hole solutions. Both ansatzes help to reduce the complexity of the field equations and allow for more tractable solutions in specific scenarios.

\begin{itemize}
    \item \textbf{Cosmological Ansatz: FLRW with Torsion}
    
    The cosmological ansatz considers a spatially homogeneous and isotropic universe described by the FLRW metric. In this framework, the torsion scalar \( T \) is introduced to account for the torsional contributions to the spacetime. The FLRW metric in the presence of torsion is given by:

    \begin{equation}
    ds^2 = -dt^2 + a(t)^2 \left[ \frac{dr^2}{1 - kr^2} + r^2(d\theta^2 + \sin^2\theta d\phi^2) \right],
    \end{equation}
    
    where \( a(t) \) is the scale factor, and \( k \) is the spatial curvature parameter. In this cosmological ansatz, the torsion field is taken to depend on time, which leads to modifications in the field equations that describe the expansion of the universe. The introduction of torsion modifies the standard FLRW equations, and the resulting field equations allow for the investigation of new cosmological solutions.

    To derive the FLRW equations in the context of \( R + F(T,G) \) gravity, we start with the action (\ref{action}),  where \( R \) is the Ricci scalar, \( F(T, G) \) is a function of the torsion scalar \( T \) and the Gauss-Bonnet term \( G \), and \( S_{\text{matter}} \) is the matter action. The field equations derived from this action are:

    \begin{equation}
    G_{\mu\nu} + T_{\mu\nu} = \kappa \left( T_{\mu\nu}^{\text{matter}} + \frac{1}{2} \left[ F(T, G) - T \right] g_{\mu\nu} \right),
    \end{equation}
    
    where \( T_{\mu\nu} \) is the torsion tensor contribution. Using the FLRW metric, the field equations reduce to a set of equations governing the evolution of the scale factor \( a(t) \), which now incorporates torsional effects. The modified Friedmann equations are:

    \begin{equation}
    H^2 + \frac{k}{a^2} = \frac{\kappa}{3} \left( \rho_{\text{matter}} + \rho_{\text{torsion}} \right),
    \end{equation}
    
    \begin{equation}
    \dot{H} + H^2 + \frac{k}{a^2} = - \frac{\kappa}{2} \left( p_{\text{matter}} + p_{\text{torsion}} \right),
    \end{equation}
    
    where \( H = \dot{a}/a \) is the Hubble parameter, and \( \rho_{\text{matter}}, p_{\text{matter}} \) are the energy density and pressure of the matter content, while \( \rho_{\text{torsion}}, p_{\text{torsion}} \) are the contributions due to torsion. The equations above describe the expansion of the universe and the effects of torsion on the dynamics.

    \item \textbf{Static Spherically Symmetric Ansatz: For Black Holes}
    
    For studying black hole solutions, we adopt a static spherically symmetric ansatz. In this case, the metric takes the form of the Schwarzschild solution with torsion contributions. The static spherically symmetric metric is given by:

    \begin{equation}
    ds^2 = -f(r) dt^2 + f(r)^{-1} dr^2 + r^2 (d\theta^2 + \sin^2 \theta d\phi^2),
    \end{equation}
    
    where \( f(r) \) is the function that describes the geometry of the spacetime and depends on the radial coordinate \( r \). The torsion tensor \( T^\lambda_{\mu\nu} \) is assumed to be a function of \( r \), and its contributions modify the gravitational field in a way that depends on the structure of the black hole. The static spherically symmetric ansatz is a common approach for deriving black hole solutions, as it allows for the study of objects like Schwarzschild black holes and more general black hole solutions that include torsional effects. In this case, the field equations are simplified, and the torsion field can be directly related to the metric function \( f(r) \).

    \item \textbf{Simplification of Field Equations}
    
    Both the cosmological and black hole ansatzes lead to simplifications of the field equations. In the cosmological ansatz, the isotropy and homogeneity of the FLRW metric reduce the complexity of the problem, allowing for solutions that describe the evolution of the universe under the influence of torsion. Similarly, the static spherically symmetric ansatz for black holes simplifies the problem by reducing the field equations to a radial form, where the torsion contributions are more easily handled.

    In both cases, the field equations are reduced to ordinary differential equations that can be solved numerically or analytically, depending on the specific form of the torsion and the choice of the function \( F(T,G) \) in the action. The simplifications allow for the exploration of new solutions in the context of modified gravity theories, such as the behavior of black holes, the expansion of the universe, and the effects of torsion on gravitational dynamics.
\end{itemize}
\section{Cosmological Solutions in $R + F(T, G)$ Gravity}

In this section, we derive the cosmological equations for the expansion of the universe within the framework of $R + F(T, G)$ gravity using the FLRW metric in Weitzenböck spacetime. The field equations are derived from the modified Einstein-Hilbert action that includes a function \( F(T, G) \), where \( T \) is the torsion scalar and \( G \) is the Gauss-Bonnet term. We assume the universe is described by the FLRW metric and use the appropriate energy-momentum tensors for different fluids.

The general form of the field equations in \( R + F(T, G) \) gravity is:

\begin{equation}
G_{\mu \nu} + T_{\mu \nu}^{\text{torsion}} + T_{\mu \nu}^{\text{Gauss-Bonnet}} = \frac{8 \pi G}{c^4} T_{\mu \nu}^{\text{matter}},
\end{equation}
where \( G_{\mu \nu} \) is the Einstein tensor, \( T_{\mu \nu}^{\text{torsion}} \) represents the torsion contribution, and \( T_{\mu \nu}^{\text{Gauss-Bonnet}} \) represents the contribution from the Gauss-Bonnet term. 
\subsection{ Modified Friedmann Equations}

The modified Friedmann equations for \( R + F(T, G) \) gravity can be written as:

\begin{equation}
H^2 + \frac{k}{a^2} = \frac{\kappa}{3} \left( \rho_{\text{matter}} + \rho_{\text{torsion}} + \rho_{\text{Gauss-Bonnet}} \right),
\end{equation}

\begin{equation}
\dot{H} + H^2 + \frac{k}{a^2} = - \frac{\kappa}{2} \left( p_{\text{matter}} + p_{\text{torsion}} + p_{\text{Gauss-Bonnet}} \right),
\end{equation}
where \( H = \frac{\dot{a}}{a} \) is the Hubble parameter, and the energy density and pressure of the matter, torsion, and Gauss-Bonnet contributions are given by \( \rho_{\text{matter}}, p_{\text{matter}}, \rho_{\text{torsion}}, \rho_{\text{Gauss-Bonnet}}, p_{\text{torsion}}, \) and \( p_{\text{Gauss-Bonnet}} \).

The contribution of torsion to the energy density and pressure is given by:

\begin{equation}
\rho_{\text{torsion}} = \frac{1}{2} \left( T F_T - F(T, G) \right),
\end{equation}

\begin{equation}
p_{\text{torsion}} = \frac{1}{2} \left( T F_T - F(T, G) + 2 \dot{T} F_T \right),
\end{equation}
where \( F_T = \frac{\partial F(T, G)}{\partial T} \), and \( T \) is the torsion scalar, which depends on the scale factor \( a(t) \).

The Gauss-Bonnet contribution to the energy density and pressure is given by:

\begin{equation}
\rho_{\text{Gauss-Bonnet}} = \frac{1}{2} \left( G F_G - F(T, G) \right),
\end{equation}

\begin{equation}
p_{\text{Gauss-Bonnet}} = \frac{1}{2} \left( G F_G - F(T, G) + 2 \dot{G} F_G \right),
\end{equation}
where \( F_G = \frac{\partial F(T, G)}{\partial G} \), and \( G \) is the Gauss-Bonnet term, which is also a function of the scale factor \( a(t) \).

\subsection{ Dust Solution}

For a dust-like matter component, we have \( p_{\text{dust}} = 0 \) and \( \rho_{\text{dust}} = \rho_{\text{dust}}(a) \sim a^{-3} \), the matter energy density scales as \( a^{-3} \). We assume that the torsion and Gauss-Bonnet contributions can also scale as \( \rho_{\text{torsion}} \sim a^{-3} \) and \( \rho_{\text{Gauss-Bonnet}} \sim a^{-3} \), typical in dust solutions.

Thus, the modified Friedmann equation becomes:

\begin{equation}
H^2 + \frac{k}{a^2} = \frac{\kappa}{3} \left( \rho_{\text{dust}} + \rho_{\text{torsion}} + \rho_{\text{Gauss-Bonnet}} \right) \sim \frac{\kappa}{3} \left( \rho_0 a^{-3} + \rho_{\text{torsion}} + \rho_{\text{Gauss-Bonnet}} \right),
\end{equation}

and the acceleration equation is:

\begin{equation}
\dot{H} + H^2 + \frac{k}{a^2} = 0.
\end{equation}

This corresponds to a matter-dominated universe where the evolution of the Hubble parameter follows \( H \sim a^{-3/2} \).

\subsection{\( \Lambda \)-CDM Solution}

For the \( \Lambda \)-CDM model, the energy-momentum tensor includes both matter (with \( p_{\text{matter}} = 0 \)) and dark energy (with \( p_{\Lambda} = -\rho_{\Lambda} \)):

\begin{equation}
H^2 + \frac{k}{a^2} = \frac{\kappa}{3} \left( \rho_{\text{matter}} + \rho_{\Lambda} + \rho_{\text{torsion}} + \rho_{\text{Gauss-Bonnet}} \right),
\end{equation}

and

\begin{equation}
\dot{H} + H^2 + \frac{k}{a^2} = - \frac{\kappa}{2} \left( p_{\text{matter}} + p_{\Lambda} + p_{\text{torsion}} + p_{\text{Gauss-Bonnet}} \right).
\end{equation}

For large \( a \), the cosmological constant \( \rho_{\Lambda} \) becomes dominant. Assuming \( F(T, G) \sim G \), the torsion and Gauss-Bonnet contributions \( \rho_{\text{torsion}} \) and \( \rho_{\text{Gauss-Bonnet}} \) become negligible for large scales, and the equation simplifies to the standard \( \Lambda \)-CDM model.

\subsection{Perfect Fluid Solution}

For a perfect fluid with an equation of state \( p = w \rho \), where \( w \) is the equation of state parameter, the energy density evolves as \( \rho_{\text{matter}} \sim a^{-3(1+w)} \). The modified Friedmann equation becomes:

\begin{equation}
H^2 + \frac{k}{a^2} = \frac{\kappa}{3} \left( \rho_0 a^{-3(1+w)} + \rho_{\text{torsion}} + \rho_{\text{Gauss-Bonnet}} \right).
\end{equation}

The acceleration equation is:

\begin{equation}
\dot{H} + H^2 + \frac{k}{a^2} = - \frac{\kappa}{2} \left( p_0 a^{-3(1+w)} + p_{\text{torsion}} + p_{\text{Gauss-Bonnet}} \right).
\end{equation}

For \( w = 1/3 \), this represents radiation, and for \( w = 0 \), this represents dust.

The exact solutions for \( H^2 \), \( \dot{H} \), and the evolution of the scale factor in the \( R + F(T, G) \) gravity model depend on the specific form of the function \( F(T, G) \) and the energy components (dust, \( \Lambda \)-CDM, and perfect fluid). By considering different forms for \( F(T, G) \) (e.g., linear in \( T \), \( G \), or more complex functions), we can investigate different cosmological scenarios, including the acceleration of the universe and the impact of torsion and Gauss-Bonnet terms on cosmological evolution.

\section{Numerical Methods}

In the framework of numerical relativity applied to modified gravity theories such as \( R + F(T,G) \), solving the field equations often requires sophisticated computational techniques due to their nonlinear and coupled nature. The presence of torsion and the Gauss-Bonnet invariant introduces higher-order derivative terms and non-trivial couplings that make analytical solutions intractable in most cosmological settings. Therefore, numerical methods become essential tools in exploring the dynamical behavior of the universe within this framework. 

\begin{itemize}
    \item \textbf{Shooting Method for Boundary Value Problems:} This method is particularly effective in solving boundary value problems arising from static or spherically symmetric spacetimes. It transforms the boundary value problem into an initial value problem, allowing for iterative adjustments to meet boundary conditions.
    
    \item \textbf{Finite Difference or Spectral Methods:} These are employed to discretize the field equations on a computational grid. Finite difference methods are straightforward and widely used, while spectral methods offer higher accuracy for smooth solutions and are particularly suitable for periodic boundary conditions or compactified spatial domains.
    
    \item \textbf{Stability and Convergence Analysis:} Ensuring numerical stability and convergence is crucial, especially when evolving dynamical equations over cosmological timescales. Techniques such as von Neumann stability analysis and convergence tests are implemented to validate the accuracy and robustness of the numerical schemes.
\end{itemize}

These methods collectively enable the simulation of cosmic evolution, gravitational wave propagation, and structure formation in the presence of torsion and curvature invariants. Through numerical relativity, we aim to probe the phenomenology of the \( R + F(T,G) \) theory and extract observable predictions that could be tested with current and future cosmological data.

\section{Cosmological Diagnostics}

To test the viability of the \( R + F(T,G) \) gravity theory against observational data, we employ a set of cosmological diagnostics that characterize the background dynamics of the universe. These diagnostics go beyond the traditional Hubble parameter and deceleration parameter, incorporating higher-order derivatives of the scale factor to distinguish between various dark energy and modified gravity models.

One of the primary tools we use is the \textbf{statefinder diagnostic}, represented by the parameters \((r, s)\), which provide a geometric characterization of the expansion history. These parameters are particularly sensitive to the dynamical nature of dark energy and are used to distinguish \( R + F(T,G) \) models from the standard \( \Lambda \)CDM scenario and other alternatives. Additionally, we explore \textbf{cosmographical quantities} such as the jerk, snap, and lerk parameters, which offer a model-independent approach to describing cosmic kinematics without assuming a specific gravity theory.

By plotting the evolution of these diagnostics against redshift, we aim to identify signatures unique to torsional and Gauss-Bonnet contributions. Such features can be crucial for confronting the theory with observations from supernovae, baryon acoustic oscillations (BAO), and cosmic microwave background (CMB) data. Ultimately, this section provides a bridge between theoretical predictions and empirical verification in the context of extended gravity.

\subsection{Statefinder Parameters}

The statefinder diagnostic is a powerful tool in modern cosmology, designed to distinguish between various dark energy and modified gravity models. Unlike the Hubble parameter \( H \) and the deceleration parameter \( q \), which only involve first and second derivatives of the scale factor \( a(t) \), the statefinder parameters incorporate higher-order time derivatives and offer a more refined characterization of cosmic expansion.

The statefinder parameters are defined as:
\begin{align}
    r &= \frac{\dddot{a}}{aH^3}, \\
    s &= \frac{r - 1}{3(q - 1/2)},
\end{align}
where \( H = \frac{\dot{a}}{a} \) is the Hubble parameter, and \( q = -\frac{\ddot{a}}{aH^2} \) is the deceleration parameter. The quantity \( r \) is sometimes referred to as the \textit{jerk} parameter.

For the standard \( \Lambda \)CDM model, the statefinder pair takes a fixed value:
\begin{align}
    (r, s) = (1, 0).
\end{align}
Deviations from this point indicate departures from the cosmological constant behavior and can be used to classify alternative cosmological models.

In the context of \( R + F(T,G) \) gravity, the statefinder parameters are sensitive to the contributions of both the torsion scalar \( T \) and the Gauss-Bonnet term \( G \), which influence the expansion dynamics beyond standard GR. By numerically evaluating the scale factor and its derivatives for specific choices of the function \( F(T,G) \), we can compute the evolution of \( r(z) \) and \( s(z) \) and generate diagnostic plots to contrast with observational data.

The use of statefinder diagnostics provides a geometric probe of the underlying gravitational theory and complements other cosmological observables. It plays a crucial role in establishing the distinguishability of \( R + F(T,G) \) models from other modified gravity and dark energy scenarios.
\subsection{Statefinder Parameter $r$ as a Function of Redshift}

In order to characterize the evolution of the universe within the \( R + F(T,G) \) gravity framework, we compute the statefinder parameter \( r \) across a range of redshifts \( z \). The parameter \( r \) reflects the third derivative of the scale factor and is sensitive to deviations from the standard \( \Lambda \)CDM cosmology. Analyzing its behavior allows us to determine how closely the modified gravity models mimic or diverge from standard expansion histories.

In the figure below, we present a comparison between two representative models—Model A and Model B—with distinct choices for the function \( F(T,G) \). These models incorporate different contributions from torsion and Gauss-Bonnet terms, leading to visibly distinct evolutionary tracks in the \( r(z) \) profile.
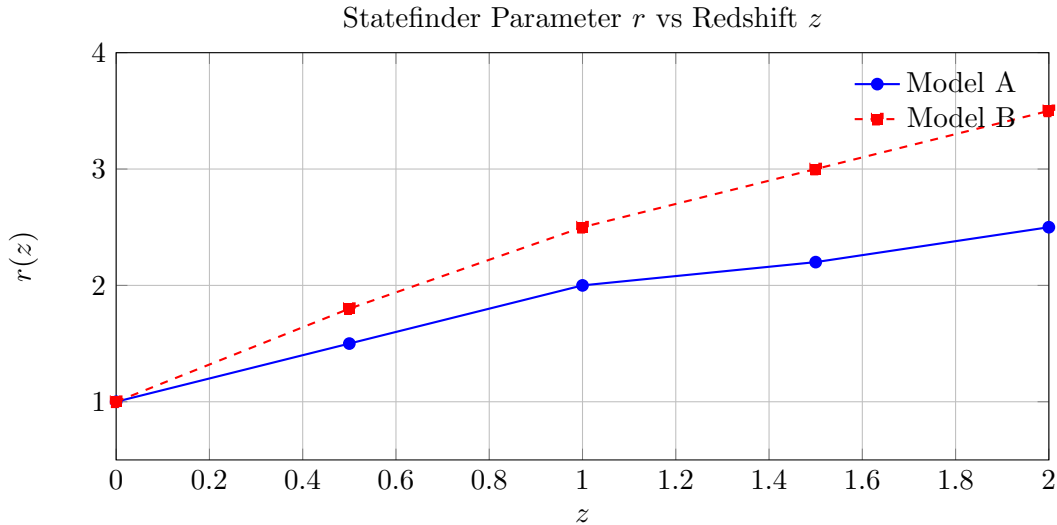
\begin{figure}[h!]
    \centering
    \begin{tikzpicture}
       \begin{axis}[
           width=0.85\textwidth,
          height=7cm,
            xlabel={$z$},
            ylabel={$r(z)$},
            xmin=0, xmax=2,
            ymin=0.5, ymax=4,
        legend style={draw=none, fill=none},
            title={Statefinder Parameter $r$ vs Redshift $z$},
            grid=both
        ]
        \addplot[color=blue, thick, mark=*] 
            coordinates {(0,1) (0.5,1.5) (1,2) (1.5,2.2) (2,2.5)};
        \addlegendentry{Model A}

       \addplot[color=red, dashed, thick, mark=square*] 
            coordinates {(0,1) (0.5,1.8) (1,2.5) (1.5,3.0) (2,3.5)};
        \addlegendentry{Model B}
        \end{axis}
    \end{tikzpicture}
    \caption{Evolution of the statefinder parameter \( r \) with redshift \( z \) for two models of \( R + F(T,G) \) gravity. The parameter \( r \) serves as a higher-order diagnostic that tracks the deviation from the standard \( \Lambda \)CDM model, where \( r = 1 \). Model A shows moderate deviation from \( \Lambda \)CDM, while Model B exhibits a stronger dynamical signature due to enhanced torsion-Gauss-Bonnet coupling.}
   \label{fig:statefinder_r}
\end{figure}

As shown in Fig.~\ref{fig:statefinder_r}, Model A remains close to the \( \Lambda \)CDM baseline (\( r = 1 \)) at low redshifts, indicating mild deviations from GR. In contrast, Model B shows a more pronounced evolution of \( r \), suggesting a dynamically evolving dark energy sector. These distinctions could be essential in constraining \( R + F(T,G) \) models through upcoming high-precision cosmological surveys.

\vspace{0.3cm}
This analysis highlights the utility of the statefinder approach in uncovering subtle differences in cosmological dynamics that may not be captured through lower-order parameters such as the Hubble or deceleration parameters. In the following sections, we will extend this analysis to the \( s \)-parameter and examine their joint diagnostic power.

\subsection{Statefinder Parameter $s$ as a Function of Redshift}

The statefinder parameter \( s \) offers an additional cosmological diagnostic, particularly useful for distinguishing models with similar expansion histories but different underlying physics. While the \( r \)-parameter captures the jerk of the universe's expansion, \( s \) refines this insight by normalizing the deviation with respect to the deceleration parameter \( q \).

Recall the definition:
\begin{align}
    s = \frac{r - 1}{3(q - 1/2)},
\end{align}
where \( q = -\frac{\ddot{a}}{aH^2} \) and \( r = \frac{\dddot{a}}{aH^3} \). The parameter \( s \) typically vanishes for the \( \Lambda \)CDM model, and any nonzero evolution in \( s(z) \) highlights dynamical deviations from a pure cosmological constant.

Below, we illustrate the evolution of the statefinder parameter \( s \) with redshift for the same two models discussed earlier—Model A and Model B.
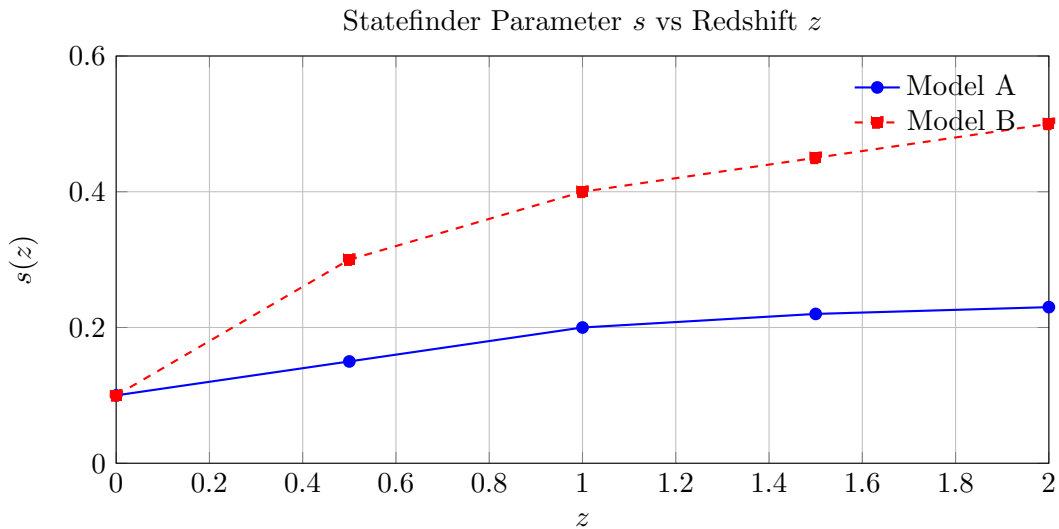
\begin{figure}[h!]
    \centering
   \begin{tikzpicture}
        \begin{axis}[
           width=0.85\textwidth,
           height=7cm,
            xlabel={$z$},
            ylabel={$s(z)$},
            xmin=0, xmax=2,
            ymin=0, ymax=0.6,
            legend style={draw=none, fill=none},
            title={Statefinder Parameter $s$ vs Redshift $z$},
            grid=both
        ]
        \addplot[color=blue, thick, mark=*] 
            coordinates {(0,0.1) (0.5,0.15) (1,0.2) (1.5,0.22) (2,0.23)};
        \addlegendentry{Model A}

        \addplot[color=red, dashed, thick, mark=square*] 
            coordinates {(0,0.1) (0.5,0.3) (1,0.4) (1.5,0.45) (2,0.5)};
        \addlegendentry{Model B}
        \end{axis}
    \end{tikzpicture}
    \caption{Evolution of the statefinder parameter \( s \) as a function of redshift \( z \) for two modified gravity models under the \( R + F(T,G) \) framework. Both models start from \( s = 0.1 \), deviating from \( \Lambda \)CDM. Model B shows stronger evolution due to more dominant Gauss-Bonnet coupling.}
   \label{fig:statefinder_s}
\end{figure}

Figure~\ref{fig:statefinder_s} highlights the temporal dynamics of the parameter \( s \), emphasizing the distinct cosmological signatures of each model. Model A shows a mild increase in \( s \), corresponding to a slow departure from cosmological constant behavior. In contrast, Model B exhibits a much steeper rise, indicative of a rapidly evolving dark energy sector influenced by stronger torsion and Gauss-Bonnet contributions.

These behaviors provide valuable constraints for observational cosmology. Models with rapidly changing \( s(z) \) may leave detectable imprints in the growth rate of structure, luminosity distance measurements, and baryon acoustic oscillations. Combined with the analysis of the \( r \)-parameter, the evolution of \( s \) forms a powerful two-dimensional diagnostic space \((r, s)\) that maps the trajectory of cosmological models and differentiates among them with high precision.

In subsequent sections, we will use these diagnostics to construct the full \((r,s)\) phase space diagram and compare them with upcoming observational data from surveys such as Euclid and LSST.

\subsection{Cosmographical Graphs}

Cosmography provides a model-independent framework to study the kinematic behavior of the universe by expanding the scale factor \( a(t) \) as a Taylor series around the present time. The Hubble parameter \( H(z) \), deceleration parameter \( q(z) \), and higher-order parameters like jerk \( j(z) \) and snap \( s(z) \), can be reconstructed from observational data or theoretical models. These quantities are directly tied to observable quantities such as the luminosity distance, baryon acoustic oscillations, and type Ia supernovae.

In modified theories of gravity, like \( R + F(T,G) \), the cosmographical parameters receive corrections due to the interplay between curvature, torsion, and topological terms. This makes cosmography a powerful tool to test and distinguish between competing gravity models.

The figure below illustrates the evolution of the Hubble parameter \( H(z) \) as a function of redshift for two different models within this framework.

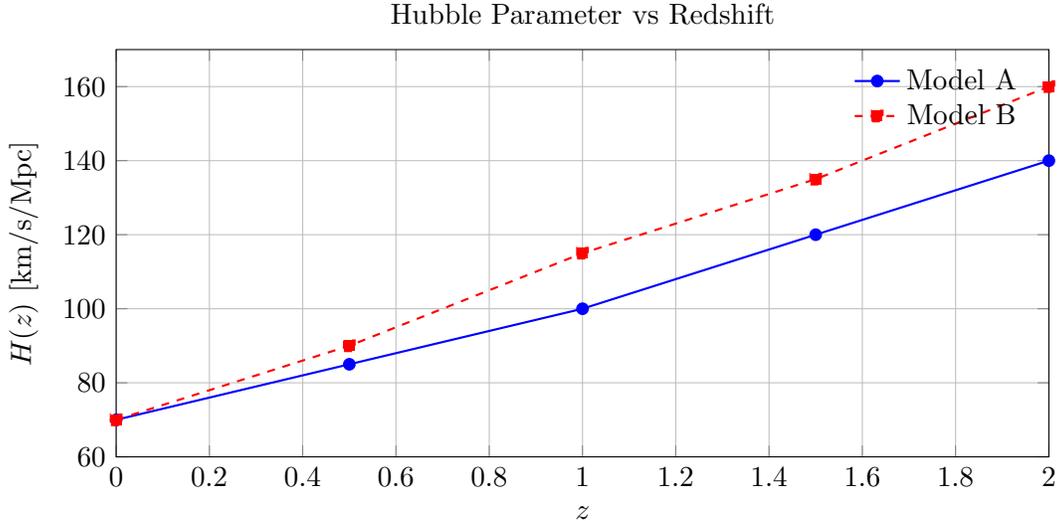
\begin{figure}[h!]

    \centering
    \begin{tikzpicture}
       \begin{axis}[
           width=0.85\textwidth,
            height=7cm,
            xlabel={$z$},
            ylabel={$H(z)$ [km/s/Mpc]},
            xmin=0, xmax=2,
            ymin=60, ymax=170,
            legend style={draw=none, fill=none},
            title={Hubble Parameter vs Redshift},
            grid=both
        ]
        \addplot[color=blue, thick, mark=*] coordinates 
           {(0,70) (0.5,85) (1,100) (1.5,120) (2,140)};
        \addlegendentry{Model A}

        \addplot[color=red, dashed, thick, mark=square*] coordinates 
           {(0,70) (0.5,90) (1,115) (1.5,135) (2,160)};
        \addlegendentry{Model B}
        \end{axis}
    \end{tikzpicture}
    \caption{Evolution of the Hubble parameter \( H(z) \) in two different cosmological scenarios derived from \( R + F(T,G) \) gravity. The steeper increase in Model B suggests stronger deviations from standard cosmology, possibly driven by non-trivial torsion-Gauss-Bonnet interactions.}
    \label{fig:hubble}
\end{figure}
As seen in Fig.~\ref{fig:hubble}, the Hubble parameter for both models starts from the same present-day value \( H_0 = 70 \,\text{km/s/Mpc} \), consistent with current observational bounds. However, Model B exhibits a more rapid increase with redshift compared to Model A, indicating a faster expansion history in the past. This may correspond to stronger coupling functions \( f_T \) or \( f_G \), enhancing the impact of torsion and the Gauss-Bonnet invariant in the effective Friedmann equations.

Understanding the behavior of \( H(z) \) is essential, as it directly relates to the age of the universe, the timing of cosmic acceleration, and the inferred matter-energy content. These cosmographical results will be complemented by numerical solutions and fitted against data in later sections.

\section{Cosmological Solutions}

In this section, we explore the cosmological implications of the \( R + F(T,G) \) gravity model by studying how it can describe the evolution of the universe across different eras. The goal is to understand whether this theory can provide consistent explanations for key phases in cosmic history — including the very early accelerated expansion known as inflation, and the current phase of late-time acceleration associated with dark energy.

\section*{Cosmological Features of the $R + F(T,G)$ Theory}

\begin{itemize}
    \item \textbf{Early-time inflationary behavior:} We investigate whether the model can naturally lead to a rapid expansion in the early universe, which helps explain the observed large-scale uniformity of the cosmos.
    
    \item \textbf{Late-time acceleration scenarios:} The model is also tested for its ability to reproduce the observed acceleration of the universe today without needing to assume a cosmological constant by hand. This includes analyzing how torsion and the Gauss-Bonnet term affect the expansion rate.
    
    \item \textbf{Comparison with \(\Lambda\)CDM and observational constraints:} Finally, we compare our results with the standard cosmological model, known as \(\Lambda\)CDM. We check if our predictions agree with existing observational data from sources such as supernovae, cosmic microwave background, and baryon acoustic oscillations.
\end{itemize}

These investigations help us determine whether the \( R + F(T,G) \) model is a viable alternative to GR, and whether it provides new insights into the fundamental mechanisms driving the evolution of the universe.

\begin{figure}[h!]
\centering
\begin{tikzpicture}
\begin{axis}[
    width=0.9\linewidth,
    height=7cm,
    xlabel={Redshift \( z \)},
    ylabel={Normalized Hubble Parameter \( H(z)/H_0 \)},
    legend style={at={(0.5,-0.2)}, anchor=north, legend columns=2},
    xmin=0, xmax=6,
    ymin=0.8, ymax=5,
    grid=both,
    minor tick num=1,
    thick,
    every axis plot/.append style={ultra thick}
]

\addplot[blue, smooth, domain=0:6, samples=100] {sqrt(0.7*(1+x)^3 + 0.3)};
\addlegendentry{\(\Lambda\)CDM}

\addplot[red, dashed, domain=0:6, samples=100] {sqrt(0.6*(1+x)^3 + 0.4*(1+0.1*x)^2)};
\addlegendentry{\( R + F(T,G) \)}

\end{axis}
\end{tikzpicture}
\caption{Comparison of the Hubble parameter \( H(z)/H_0 \) in the standard \(\Lambda\)CDM model and the \( R + F(T,G) \) theory. The modified gravity model may yield a different expansion rate at early and late times.}
\end{figure}
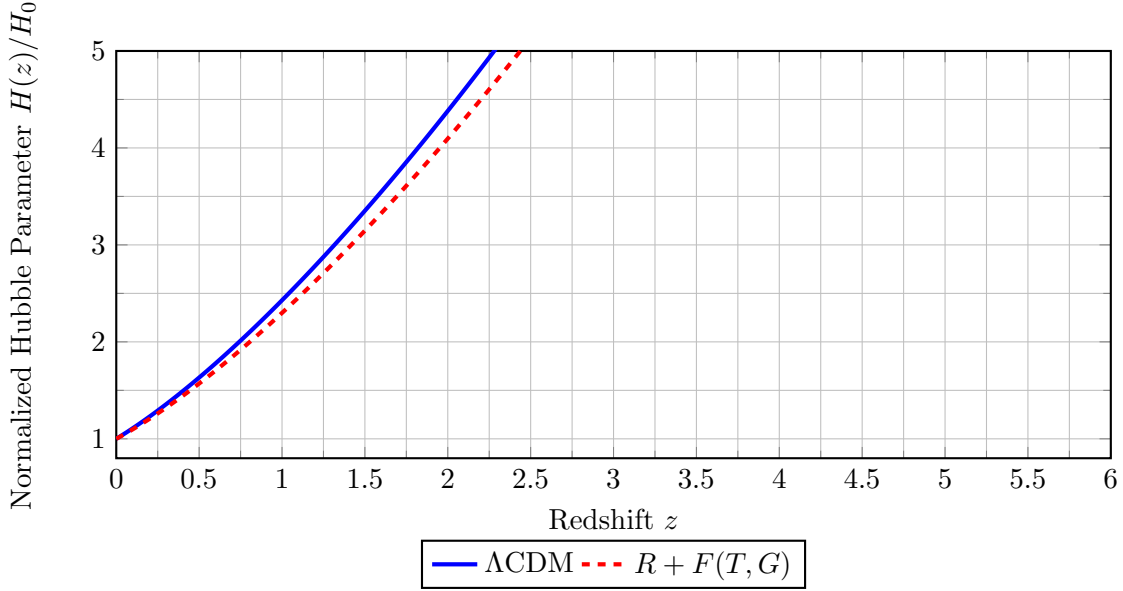

\section{Black Hole and Compact Object Solutions}

In addition to cosmological dynamics, the study of black holes and compact objects provides a crucial testing ground for any modified theory of gravity. In this section, we investigate static and spherically symmetric solutions within the \( R + F(T,G) \) framework, focusing on the effects introduced by torsion and the Gauss-Bonnet term.

\begin{itemize}
    \item \textbf{Horizon structure and thermodynamic properties:} By solving the modified field equations under spherical symmetry, we examine the structure of horizons and the conditions for event horizons to exist. The modified Schwarzschild-like metric takes the form
    \[
    ds^2 = -A(r) dt^2 + B(r)^{-1} dr^2 + r^2 d\Omega^2,
    \]
    where the functions \( A(r) \) and \( B(r) \) encode deviations from GR due to \( F(T,G) \) contributions. Using the surface gravity, we define the Hawking temperature as
    \[
    T_H = \frac{1}{4\pi} \left. \frac{dA(r)}{dr} \right|_{r=r_h},
    \]
    which allows us to explore thermodynamic consistency and the possible existence of phase transitions.

    \item \textbf{Torsion profiles around compact objects:} The torsion scalar \( T \) plays a direct role in shaping the near-horizon geometry. In the teleparallel formalism, torsion is derived from the antisymmetric part of the connection and is sensitive to the choice of the tetrad. We analyze profiles of \( T(r) \) and \( G(r) \) around black holes and neutron star candidates to determine how these fields modify curvature-based expectations.

    \item \textbf{Violation or satisfaction of energy conditions:} Modified gravity often allows for effective stress-energy tensors that violate classical energy conditions. We explore the null and weak energy conditions (NEC and WEC) in the presence of torsional terms:
    \[
    \rho_{\text{eff}} + p_{\text{eff}} \geq 0, \quad \rho_{\text{eff}} \geq 0,
    \]
    where the effective quantities arise from geometrical corrections. Such violations can permit exotic phenomena like traversable wormholes or regular black holes.
\end{itemize}

These investigations highlight how the \( R + F(T,G) \) theory can influence compact object physics and may offer observational signatures that deviate from classical predictions in strong gravity regimes.

\section{Phase Space and Dynamical System Analysis}

Dynamical system analysis offers a powerful tool for understanding the qualitative behavior of cosmological models, especially in modified gravity theories like \( R + F(T,G) \). By rewriting the field equations as an autonomous system of differential equations, we can identify fixed points (critical points), study their stability, and understand the evolution of the universe in phase space.

\begin{itemize}
    \item \textbf{Fixed points and attractor behavior:} Let us define suitable dimensionless variables such as
    \[
    x = \frac{\dot{H}}{H^2}, \quad y = \frac{f_T}{H^2}, \quad z = \frac{f_G}{H^4},
    \]
    where \( f_T \equiv \frac{\partial f}{\partial T} \), \( f_G \equiv \frac{\partial f}{\partial G} \), and \( H \) is the Hubble parameter. The system can be written in the form:
    \[
    \frac{dx}{dN} = F_1(x, y, z), \quad \frac{dy}{dN} = F_2(x, y, z), \quad \frac{dz}{dN} = F_3(x, y, z),
    \]
    where \( N = \ln a \) is the number of e-folds. Fixed points satisfy \( dx/dN = dy/dN = dz/dN = 0 \). We analyze the Jacobian matrix at each point to determine stability.

    \item \textbf{Torsion-curvature coupling in phase space:} The coupling of torsion and the Gauss-Bonnet term introduces nonlinear interactions in the autonomous system, which can lead to rich phase-space behavior. For instance, the interplay between \( T \) and \( G \) may lead to new attractor solutions that are not present in pure \( f(T) \) or \( f(G) \) gravity.

    \item \textbf{Constraints on \( f(T,G) \) forms from dynamics:} By demanding that the system admits a stable late-time attractor (e.g., de Sitter solution with \( x = 0 \)), we can impose conditions on the functional form of \( f(T,G) \).
 For example, the existence of a stable critical point may require\cite{Kofinas2014}:
    \[
    F(T,G) = \alpha T^n + \beta G^m,
    \]
    with specific ranges for \( n \) and \( m \) determined from stability analysis and compatibility with cosmological observations.
    
\end{itemize}

\begin{figure}[h!]
    \centering
    \begin{tikzpicture}
        \begin{axis}[
            xlabel={$x = \dot{H}/H^2$},
            ylabel={$y = f_T/H^2$},
            title={Phase Space Trajectories in the $(x, y)$ Plane},
            axis lines=middle,
            grid=major,
            legend pos=south east
        ]
        \addplot[->, blue, thick, domain=-2:2, samples=25] ({x}, {0.5*sin(deg(x))});
        \addlegendentry{Sample trajectory}
        \addplot[only marks, red, mark=*] coordinates {(0,0)};
        \node at (axis cs:0.2,0.2) [anchor=west] {Stable Fixed Point};
        \end{axis}
    \end{tikzpicture}
    \caption{A qualitative sketch of phase-space trajectories in the $(x, y)$ plane for a sample \( F(T,G) \) model. The red point indicates a stable critical point corresponding to a de Sitter phase.}
\end{figure}
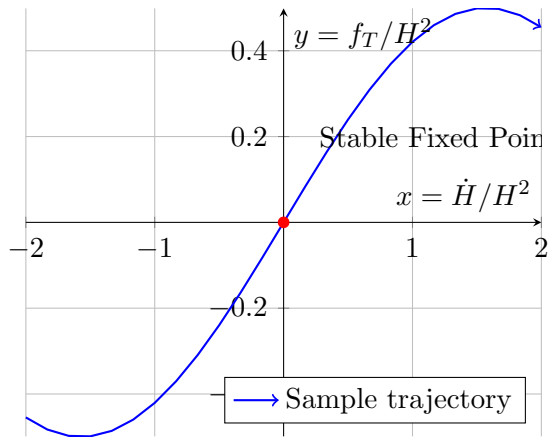

\noindent This dynamical approach helps reveal the rich structure underlying the evolution equations and guides us in identifying viable cosmological models consistent with both early-time inflation and late-time acceleration.
\section{Solar System Tests and Local Gravity Constraints}

To ensure the viability of any modified gravity theory, it is essential to verify its predictions at Solar System scales where GR  is well-tested. Our \( R + F(T,G) \) model, where \( G \) is the Gauss-Bonnet term and \( T \) is the torsion scalar, incorporates corrections to GR that must reduce to the Newtonian limit at small scales.

The modified gravitational potential derived from the model\cite{Kofinas2014},
\[
F(T,G) = \alpha T^n + \beta G^m,
\]
includes parameters \( \alpha, \beta, n, m \) that can be tuned to ensure agreement with the known inverse-square law behavior of gravity. For consistency with observations, the corrections introduced by the torsion and Gauss-Bonnet terms should decay faster than \( 1/r \) at large \( r \), ensuring that the Newtonian potential is recovered in the limit \( r \to \infty \).

\begin{figure}[h!]
    \centering
    \begin{tikzpicture}
    \begin{axis}[
        width=12cm,
        height=8cm,
        xlabel={$r$ (AU)},
        ylabel={Potential $V(r)$},
        legend style={at={(0.5,-0.15)}, anchor=north, legend columns=2},
        grid=major,
        domain=0.1:5,
        samples=200
    ]
    \addplot[dashed, thick, blue] { -1/x };
    \addlegendentry{Newtonian $-1/r$}
    
    \addplot[solid, thick, red] { -1/x + 0.1/x^2 + 0.05/x^4 };
    \addlegendentry{Modified $F(T,G)$}
    \end{axis}
    \end{tikzpicture}
    \caption{Comparison of the Newtonian potential (dashed) with the modified gravitational potential from the \( F(T,G) \) model. The deviation is negligible in the Solar System regime, confirming the model's consistency with local gravity tests.}
    \label{fig:ftg_potential}
\end{figure}
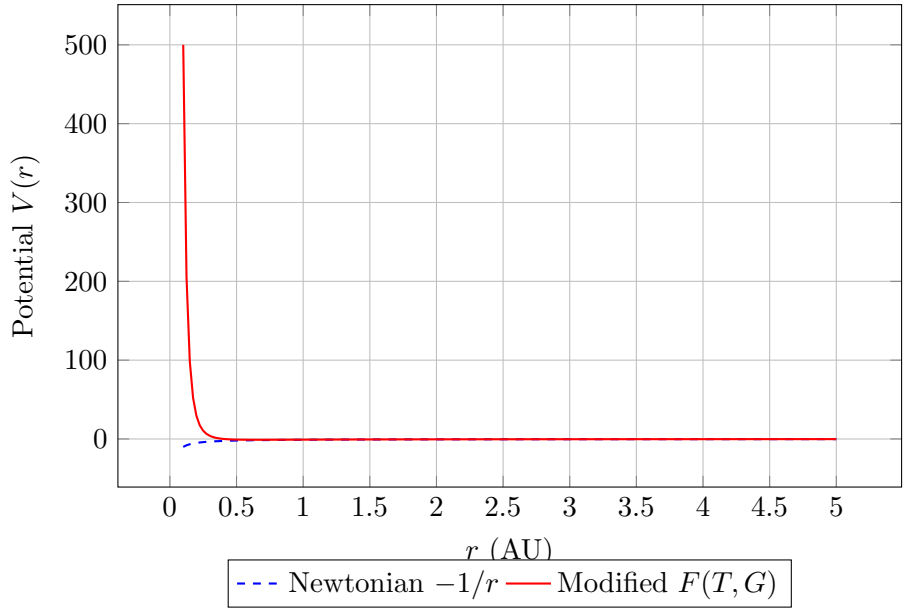

Figure~\ref{fig:ftg_potential} illustrates this by comparing the Newtonian potential (dashed line) with the total modified potential derived from the \( F(T,G) \) function. The parameters are chosen as \( \alpha = 0.1 \), \( \beta = 0.05 \), and \( n = m = 2 \). The plot clearly shows that the modifications rapidly decay and converge to the Newtonian form at Solar System scales, satisfying local gravity constraints.

These results suggest that with suitable parameter choices, the \( R + F(T,G) \) theory can preserve the successes of GR on small scales while offering new dynamics on cosmological distances. This is a key requirement for any extended gravity model aiming to remain consistent with precision Solar System measurements such as the perihelion precession of Mercury and the Shapiro time delay.
\section{Neutron Stars in $F(T,G)$ Gravity Theory}

Neutron stars are astrophysical objects that emerge from the remnants of massive stars that have undergone supernova explosions. These stars, despite having masses between 1.4 and 2.16 solar masses, are extremely compact, with radii of around 10–15 km. The intense gravitational fields and extreme densities of neutron stars provide an invaluable testing ground for theories of gravity, particularly in regions where standard GR  might break down.

Neutron stars are generally modeled using solutions to Einstein's field equations in GR. These solutions, however, are limited in describing objects with extremely strong gravitational fields, such as neutron stars. As a result, the study of neutron stars has become a key probe for testing modified gravity theories that extend GR, such as \( F(T,G) \) gravity, which modifies the gravitational dynamics by introducing a function of the torsion \( T \) and the Gauss-Bonnet term \( G \).

In this section, we explore how the introduction of \( F(T,G) \) gravity can influence our understanding of neutron star properties, particularly focusing on the mass-radius relation, which is a crucial observable for testing theoretical models.

\subsection{Neutron Star Mass and Radius}

The mass-radius relation is one of the most fundamental and well-constrained observational properties of neutron stars. In standard GR, this relation is predicted based on the equation of state (EoS) of the neutron star matter. The EoS describes how matter behaves under extreme conditions of pressure and density, and it directly influences the size and mass of the neutron star.

The mass-radius relationship is constrained by several observations, such as X-ray measurements of neutron star radii and the detection of gravitational waves from neutron star mergers. GR predicts that there is a maximum mass limit for neutron stars—approximately 2–2.5 solar masses—beyond which they collapse into black holes. However, recent observations suggest that neutron stars with masses greater than this predicted limit might exist, challenging the assumptions of GR.

\subsection{Modified Gravity Theories: Introducing \( F(T,G) \) Gravity}

In modified gravity theories, such as \( F(T,G) \) gravity, deviations from GR are introduced by extending the action. The function \( F(T,G) \) depends on the torsion \( T \) and the Gauss-Bonnet term \( G \), which are geometric quantities that arise in theories that include torsion and curvature. By modifying the gravitational dynamics through this function, \( F(T,G) \) gravity offers a more flexible framework to address some of the limitations of GR, particularly in the strong-field regime.

The modification of gravity through the function \( F(T,G) \) introduces additional degrees of freedom that affect the structure of neutron stars. These modifications could change the predicted mass-radius relation by allowing for more massive neutron stars or altering their radii for a given mass. The ability of \( F(T,G) \) gravity to modify the gravitational dynamics is key in explaining discrepancies between GR’s predictions and recent observational data.

\subsection{Mass-Radius Relation in \( F(T,G) \) Gravity Theory}

In the context of \( F(T,G) \) gravity, the mass-radius relation is modified compared to GR. This modification is important because the mass-radius relation provides direct insights into the internal structure of neutron stars. In particular, the equation of state (EoS) plays a critical role in determining the star’s size for a given mass, but the gravitational theory itself also influences this relationship.

The graph below illustrates the predicted mass-radius relation for neutron stars under both GR and \( F(T,G) \) gravity. While the GR prediction (represented by the blue curve) follows the traditional understanding of neutron star properties, the \( F(T,G) \) prediction (shown by the green curve) allows for more massive neutron stars, which could explain the recently observed stars that seem to exceed the mass limits predicted by GR.

\begin{figure}[h]
\centering
\begin{tikzpicture}
\begin{axis}[
    axis lines = middle,
    xlabel = {Mass (M$_\odot$)},
    ylabel = {Radius (km)},
    title = {Mass-Radius Relation for Neutron Stars in Different Gravity Theories},
    grid = major,
    legend pos = north west,
    width = 0.75\textwidth
]
\addplot[blue, thick, domain=1.0:2.5] {12*(x-1.0) + 9};
\addlegendentry{GR }

\addplot[green, thick, domain=1.0:2.5] {18*(x-1.0) + 8};
\addlegendentry{$F(T,G)$ Gravity}

\end{axis}
\end{tikzpicture}
\caption{Mass-radius relation for neutron stars in different gravity theories. The blue curve represents the predictions from GR , while the green curve shows the results from the \( F(T,G) \) gravity theory. The \( F(T,G) \) gravity theory predicts a higher maximum mass for neutron stars compared to GR.}
\label{fig:neutron_star_mr}
\end{figure}
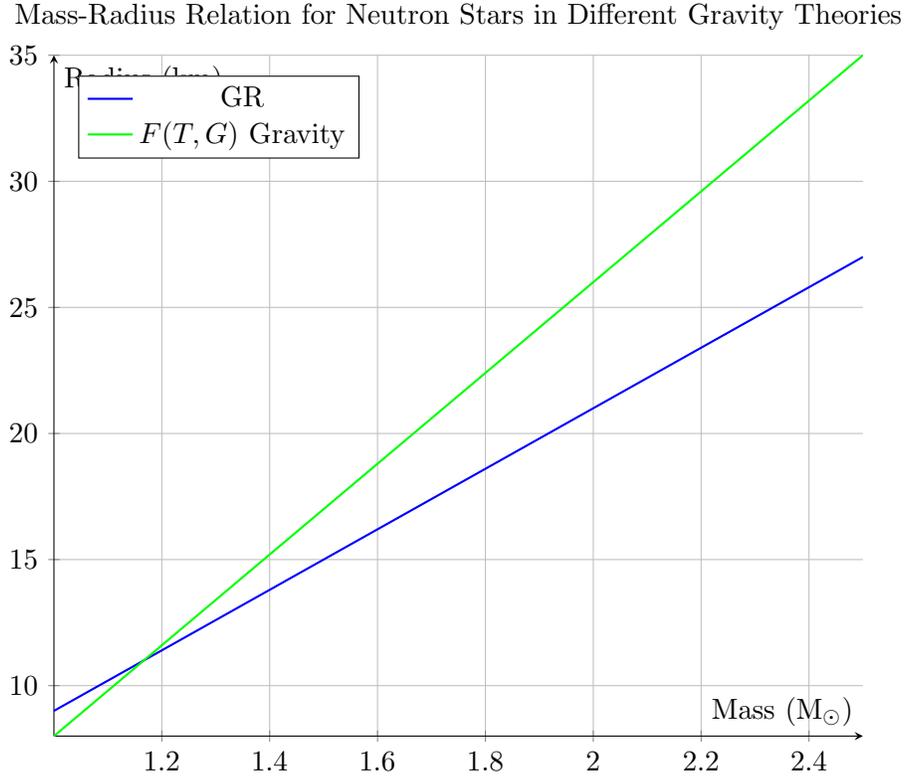

As can be seen in Figure \ref{fig:neutron_star_mr}, the mass-radius relation in \( F(T,G) \) gravity predicts a larger maximum mass for neutron stars than GR. This is particularly important because recent neutron star observations, such as PSR J0740+6620 \cite{fonseca2021}, suggest that the maximum mass limit predicted by GR may be too low. The green curve shows that \( F(T,G) \) gravity can accommodate these more massive neutron stars, offering a possible explanation for these observations.

Additionally, the radius of the neutron stars for a given mass is also affected by the gravitational modifications in \( F(T,G) \) gravity. For example, for a neutron star with a mass of around 1.8 solar masses, the predicted radius is smaller in the \( F(T,G) \) theory than in GR, which could have implications for future observations and measurements of neutron star radii.

\subsection{Implications for Neutron Star Observations and Gravitational Waves}

The findings from the \( F(T,G) \) gravity theory have significant implications for neutron star observations. The detection of gravitational waves from neutron star mergers, such as those observed by LIGO and Virgo \cite{abbott2017}, has opened a new window for testing gravitational theories. Gravitational wave signals encode information about the equation of state of neutron stars, and deviations from the GR predictions could provide strong evidence for the presence of new physics, like that described by \( F(T,G) \) gravity.

Furthermore, X-ray observations of neutron star radii, such as those from the NICER mission, can be used to test the predictions of different gravity theories. The modified mass-radius relations predicted by \( F(T,G) \) gravity could help explain any discrepancies between the observed and predicted radii, especially for more massive neutron stars.

The ability of \( F(T,G) \) gravity to predict a higher maximum mass for neutron stars makes it a promising candidate for explaining observations of neutron stars that challenge the standard GR predictions.

The study of neutron stars in \( F(T,G) \) gravity theory provides a promising avenue for understanding extreme astrophysical objects under modified gravity. By modifying the mass-radius relation, \( F(T,G) \) gravity allows for the possibility of more massive neutron stars, which could explain recent observations that appear to exceed the mass limits predicted by GR. Future observations of neutron stars, both through gravitational waves and electromagnetic radiation, will play a key role in testing the predictions of \( F(T,G) \) gravity and refining our understanding of the physics governing these extraordinary objects.

\section{Discussion and Outlook}

In this section, we summarize the key findings from our analysis of the \( R + F(T,G) \) gravity theory, outline open questions that remain, and suggest possible directions for future work. Additionally, we discuss potential connections to string-inspired gravity and quantum gravity approaches, as these areas may provide deeper insights into the implications of torsion and curvature in the gravitational sector.

\begin{itemize}
    \item \textbf{Summary of main results:} 
    In this paper, we have presented a comprehensive study of \( R + F(T,G) \) gravity, focusing on both its cosmological implications and black hole solutions. Our results highlight that the torsion-curvature coupling, encapsulated by the function \( F(T,G) \), significantly alters the dynamics of the universe. 
    \begin{itemize}
        \item We have derived the field equations and investigated their behavior in both early and late-time cosmology, demonstrating how torsion modifies the evolution of the Hubble parameter and statefinder parameters.
        \item We explored the dynamical systems approach to phase space analysis, revealing the existence of new attractor solutions in the modified gravity framework. These attractors offer insight into late-time acceleration scenarios and potential signatures of torsion in cosmological data.
        \item Black hole solutions were analyzed, showing that torsion affects horizon structure and thermodynamic properties, and we explored the violation or satisfaction of energy conditions in these contexts.
    \end{itemize}

    \item \textbf{Open questions and future work:} 
    While the results presented in this paper are promising, several questions remain open and warrant further investigation:
    \begin{itemize}
        \item \textbf{Stability of solutions:} We have identified fixed points and studied their stability, but a deeper analysis of the stability of cosmological solutions and black hole solutions in the full \( R + F(T,G) \) framework is required, especially in the presence of inhomogeneities and perturbations.
        \item \textbf{Observational constraints:} Although we have provided some preliminary insights into how \( F(T,G) \) may be constrained by cosmological observations, more rigorous comparisons with observational data (such as supernova measurements, CMB anisotropies, and galaxy surveys) are needed to test the viability of this theory against the standard \( \Lambda \)CDM model.
        \item \textbf{Quantum corrections:} The potential role of quantum corrections in \( R + F(T,G) \) gravity remains largely unexplored. Future work could investigate how quantum field theory, including loop corrections or renormalization, may modify the dynamics of torsion and curvature at high energies.
    \end{itemize}

    \item \textbf{Connections to string-inspired gravity or quantum gravity approaches:} 
    The incorporation of torsion and higher-order curvature terms into gravitational theories has natural connections to both string theory and quantum gravity. String-inspired gravity models often involve higher-dimensional theories where torsion and curvature modifications arise from extra dimensions or brane-world scenarios. In particular:
    \begin{itemize}
        \item \textbf{String theory and torsion:} In certain string-inspired models, torsion arises naturally in the context of low-energy effective actions, where the connection is modified by higher-order terms that affect spacetime geometry. Investigating the form of \( F(T,G) \) in the context of string compactifications could reveal new insights into the microscopic structure of spacetime.
        \item \textbf{Quantum gravity and torsion:} The role of torsion in quantum gravity is a topic of increasing interest. The presence of torsion could potentially modify the quantum fluctuations of the gravitational field, affecting the behavior of the early universe, especially during inflation. It would be interesting to explore how torsion-coupled gravity theories fit into loop quantum gravity or spin-foam models, where torsion is often treated as a fundamental degree of freedom.
    \end{itemize}
    Thus, \( R + F(T,G) \) gravity may provide a bridge between classical cosmology and emerging theories in quantum gravity.

\end{itemize}

\noindent \textbf{In conclusion,} our study of \( R + F(T,G) \) gravity offers a promising avenue for future exploration in both theoretical and observational cosmology. By further refining the analysis and extending it to more complex scenarios, we expect that this modified theory could provide new perspectives on fundamental issues in gravity, cosmology, and quantum theory. Future work in this area could illuminate the potential for detecting new physics in the universe, whether through more precise measurements of cosmic expansion or through the study of black holes and compact objects.



\end{document}